# Geodesics in a Toroidal space-time

S.B.P. Wickramasuriya[*], V. Joseph[#], & K.I.S. Karunaratne[+]

Abstract: We take a three dimensional Euclidean metric in toroidal coordinates and consider the corresponding Laplace equation. The simplest solution of this equation is taken. Based on this we build a Weyl space-time. This space-time is transformed to cylindrical coordinates. It is shown by using 'Mathematica' that Weyl equations in cylindrical coordinates are satisfied. Geodesic motion is considered along the symmetric axis as well as along the radii of the singularity, which is the cause of the space time.

Here we present a vacuum space-time, in cylindrical coordinates due to a ring shaped singularity. It satisfies Weyl's equations [1], [2] for axially symmetric metrics. It was originally found with the aid of a toroidal metric, obtained by the conformal transformation [3],

$z + i\rho = a \cot\left[\frac{1}{2}(\psi + i\sigma)\right]$. Then by a coordinate transformation it is brought into Weyl's form. This method of construction will be described in a paper to be published in the near future [4]. The potential $U$ satisfying Laplace's Equation in toroidal coordinates has zero value in the disc enclosed by the ring.

The metric in toroidal form is
$$ds^2 = e^{2U} dt^2 - e^{2\lambda + 2\upsilon - 2U}(d\sigma^2 + d\psi^2) - e^{2\mu - 2U} d\phi^2$$

Where
$$U = (\cosh\sigma - \cos\psi)^{\frac{1}{2}} \cos(\psi/2)$$
$$\lambda = -(1/2)\sinh^2(\sigma/2)\left(1 + \cosh^2(\sigma/2)\cos(2\psi)\right)$$
$$\upsilon = \log\left[a/(\cosh\sigma - \cos\psi)\right]$$
$$\mu = \upsilon + \log(\sinh\sigma))$$

This is one of Weyl's axially symmetric metrics in a different guise. By the coordinate transformation

$$\psi = \text{ArcCos}\left[\frac{(r^2 + z^2 - a^2)}{\left(\sqrt{(r-a)^2 + z^2}\sqrt{(r+a)^2 + z^2}\right)}\right]$$

$$\sigma = \text{ArcCosh}\left[\frac{(r^2 + z^2 + a^2)}{\left(\sqrt{(r-a)^2 + z^2}\sqrt{(r+a)^2 + z^2}\right)}\right]$$

$a$ is a constant.

[*] Department of Mathematics, University of Kelaniya, Sri Lanka. , E-mail : wickram@kln.ac.lk, wickram1930@yahoo.com
[#] 127/1, Ananda Rajakaruna Mawatha, Colombo 10, Sri Lanka.
[+] Department of Mathematics, University of Kelaniya, Sri Lanka. , E-mail : sidanthak@yahoo.com

This can be converted in to one of Weyl's cylindrically symmetric metrics

$$ds^2 = e^{2U} dt^2 - e^{2\lambda - 2U}(dr^2 + dz^2) - e^{-2U} r^2 d\phi^2$$

Then

$$\lambda = -\frac{1}{4}\left(-1 + \frac{a^2 + r^2 + z^2}{\sqrt{(a-r)^2 + z^2}\sqrt{(a+r)^2 + z^2}}\right)$$

$$\left(1 + \frac{1}{2}\left(-1 + \frac{2(-a^2 + r^2 + z^2)^2}{((a-r)^2 + z^2)((a+r)^2 + z^2)}\right)\right)\left(1 + \frac{a^2 + r^2 + z^2}{\sqrt{(a-r)^2 + z^2}\sqrt{(a+r)^2 + z^2}}\right)$$

$$U = -\sqrt{\frac{a^2}{\sqrt{(a-r)^2 + z^2}\sqrt{(a+r)^2 + z^2}}}\sqrt{1 + \frac{-a^2 + r^2 + z^2}{\sqrt{(a-r)^2 + z^2}\sqrt{(a+r)^2 + z^2}}}$$

$$\mu = \log[r]$$

By using 'Mathematica' we can show that the following Weyl's equations are satisfied.

$$\frac{\partial^2 U}{\partial r^2} + \frac{1}{r}\frac{\partial U}{\partial r} + \frac{\partial^2 U}{\partial z^2} = 0$$

$$\frac{1}{r}\frac{\partial \lambda}{\partial r} = \left(\frac{\partial U}{\partial r}\right)^2 - \left(\frac{\partial U}{\partial z}\right)^2$$

$$\frac{1}{r}\frac{\partial \lambda}{\partial z} = 2\frac{\partial U}{\partial r}\frac{\partial U}{\partial z}$$

Thus the metric is seen to assume the form

$$ds^2 = e^{2U} dt^2 - e^{-2U + 2\lambda}(dr^2 + dz^2) - e^{-2U} r^2 d\phi^2, \ U = U(r,z), \ \lambda = \lambda(r,z)$$

when transformed to that of Weyl.

This metric has a singularity on the ring $r = a$, $z = 0$.

A coordinate free definition for the speed of a test particle will also be given. Difficulties arose when trying to express the speed in terms of coordinates. For example when attempting to define speed along the radii one had to try several definitions such as $\frac{dr}{dt}, \frac{dr}{ds}, \frac{e^{-U} dr}{dt}, \frac{e^{-U} dr}{e^U dt}, \frac{e^{-U} dr}{ds}$.

Speed will be defined as $\tanh \vartheta$ where $\cosh \vartheta$ is the scalar product of the 4-velocity of the test particle with the 4-velocity of an observer stationed at the point through which the test particle is passing. Let the coordinates of the observer at rest be $(t, r, 0, 0)$. Since $r$ is constant the 4-velocity of the observer is given by $u^\mu = (u^0, 0, 0, 0)$. The four-velocity of the test particle is $v^\mu = (v^0, v^1, 0, 0)$, if it is in motion along a radius while $v^\mu = (v^0, 0, v^2, 0)$ if it is in motion along the symmetry axis $(r = 0, \phi = const)$.

$\cosh \vartheta = g_{\mu\nu} u^\mu v^\nu = g_{00} u^0 v^0$ (in both cases). We have to find $u^0$ as well as $v^0$. Obviously, $u^0$ can be found from $g_{\mu\nu} u^\mu u^\nu = 1$, and it is seen that $u^0 = e^{-U}$. To find $v^0 = \dfrac{dt}{ds}$, we use the abbreviated Lagrangian $L = e^{2U} \dot{t}^2 - e^{-2U+2\lambda} \dot{r}^2$, where the overdot represents differentiation with respect to $s$.

$\dfrac{d}{ds} \dfrac{\partial L}{\partial \dot{t}} - \dfrac{\partial L}{\partial t} = 0$ gives $\dfrac{\partial L}{\partial \dot{t}} = const$, because $\dfrac{\partial L}{\partial t} = 0$ (being a static solution).

Therefore $e^{2U} \dot{t} = E$ (say). Here $E$ is a constant related to the initial energy of the test particle.

$v^0 = \dfrac{dt}{ds} = E e^{-2U}$

$\cosh \vartheta = g_{00} u^0 v^0 = e^{2U} (e^{-U})(E e^{-2U}) = e^{-U} E$

Speed $= \tanh \vartheta = \sqrt{1 - E^{-2} e^{2U}}$

To study some of the properties of the space-time we investigate the geodesic motion along the symmetric axis normal to the ring and along the radii emanating from the centre of the ring.

For the purpose of obtaining the geodesics, we take the Lagrangian to be
$L = e^{2U} \dot{t}^2 - e^{2\lambda - 2U} (\dot{r}^2 + \dot{z}^2) - e^{-2U} r^2 \dot{\phi}^2$

Then the equations for the geodesics are

$\dfrac{d}{ds} \dfrac{\partial L}{\partial \dot{t}} - \dfrac{\partial L}{\partial t} = 0, \quad \dfrac{d}{ds} \dfrac{\partial L}{\partial \dot{r}} - \dfrac{\partial L}{\partial r} = 0, \quad \dfrac{d}{ds} \dfrac{\partial L}{\partial \dot{z}} - \dfrac{\partial L}{\partial z} = 0, \quad \dfrac{d}{ds} \dfrac{\partial L}{\partial \dot{\phi}} - \dfrac{\partial L}{\partial \phi} = 0$

We use only the first which gives us

$\dfrac{d}{ds} 2 e^{2U} \dot{t} = 0, \quad e^{2U} \dfrac{dt}{ds} = e^{2U} \dot{t} = E \, (const).$

The constant $E$ is related to the energy of the particle.

If we keep $\phi =$ constant, $z = 0$ then,
we get geodesic motion along a radius of the ring.

$ds^2 = e^{2U} dt^2 - e^{2\lambda - 2U} dr^2$

$(e^{2U} E^{-1})^2 dt^2 = e^{2U} dt^2 - e^{2\lambda - 2U} dr^2$

$e^{2\lambda - 2U} dr^2 = (e^{2U} - e^{4U} E^{-2}) dt^2$

$$e^{\lambda-U} dr = \pm e^U \left(1 - e^{2U} E^{-2}\right)^{1/2} dt$$

$$dt = -\frac{e^{\lambda-U}}{e^U \left(1 - e^{2U} E^{-2}\right)^{1/2}} dr$$

The geodesics along the radii $(z = 0, \phi = const)$ are given by

$$-e^{\lambda-U} dr = e^U \left(1 - E^{-2} e^{2U}\right)^{1/2} dt$$

Integrating,

$$t = -\int^r \frac{e^{\lambda-U} dr}{e^U \left(1 - e^{2U} E^{-2}\right)^{1/2}}$$

Here $\lambda$, $U$ are functions of $r$ only.

For geodesic motion along the symmetry axis $(r = 0)$,
we have similar equations.

$$-e^{\lambda-U} dz = e^U \left(1 - E^{-2} e^{2U}\right)^{1/2} dt$$

$$t = -\int^z \frac{e^{\lambda-U} dz}{e^U \left(1 - e^{2U} E^{-2}\right)^{1/2}}$$

with $\lambda, U$ being functions of $z$ only.

These two integrals give us two space-time diagrams involving *(t, r)* and *(t, z)*. In what follows the radius *a* is taken to be unity.

### Speed along the radial geodesic and the *(t, r)* curve

When $a = 1$, radial speeds are given by

$$\left(1 - E^{-2} e^{2U}\right)^{1/2}, \quad U = U(r).$$

This shows that by the time the rest particle reaches the periphery of the ring, its speed approaches that of the velocity of light.

For different values of *E*=1, 2 and 4, ( $a = 1$ ), we get:

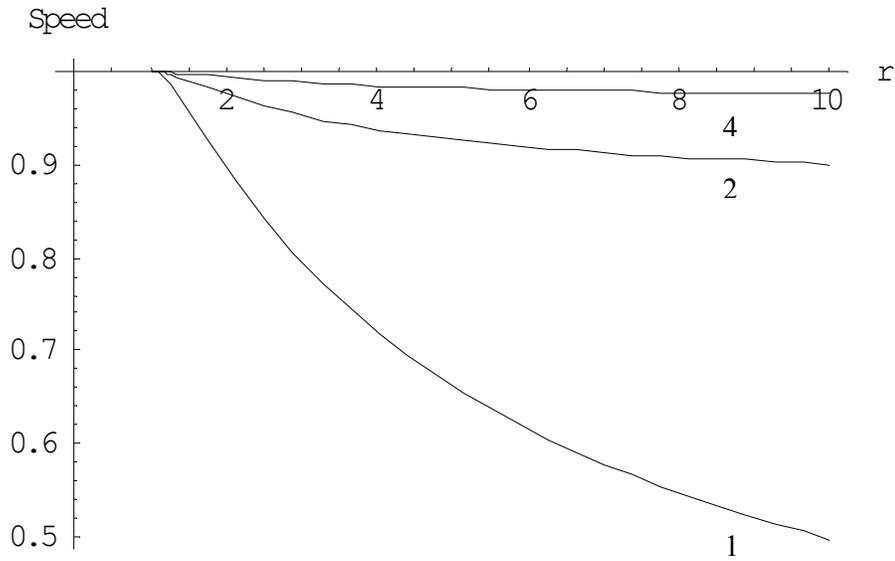

Graph of Speed versus radial distance *r*, from *r* =10 to 1 along radial geodesics for values of *E* = 1, 2 and 4, assuming that the test particles are travelling from infinity.

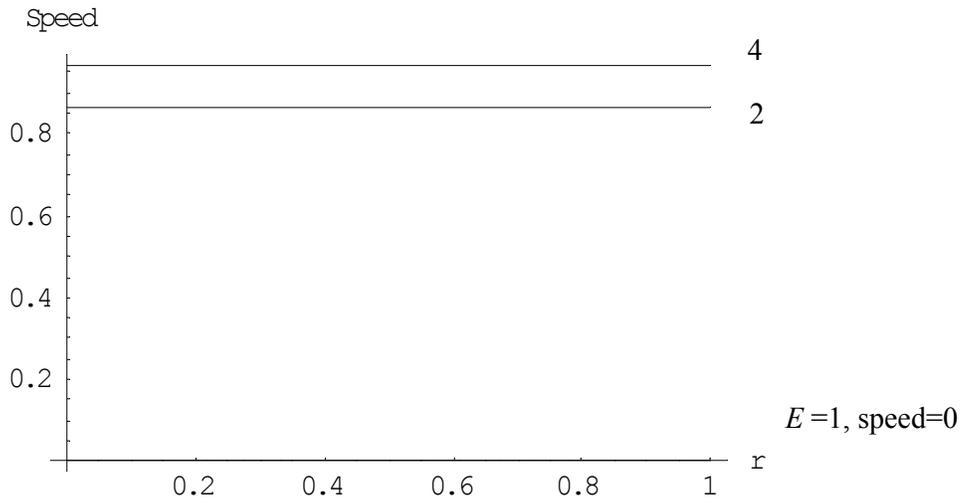

Graph of Speed versus radial distance *r*, from *r* = 0 to the periphery of the ring for *a* = 1, *E* = 1, 2 and 4. For *E* = 1 the speed is zero. As can be seen the speed remain constant, because U = 0.

When $r = 10$, $t = 0$ solving $t'[r] = -\dfrac{e^{\lambda-2U}}{\sqrt{1-E^{-2}e^{2U}}}$

Integrating with respect to r from 10 to 1, we get

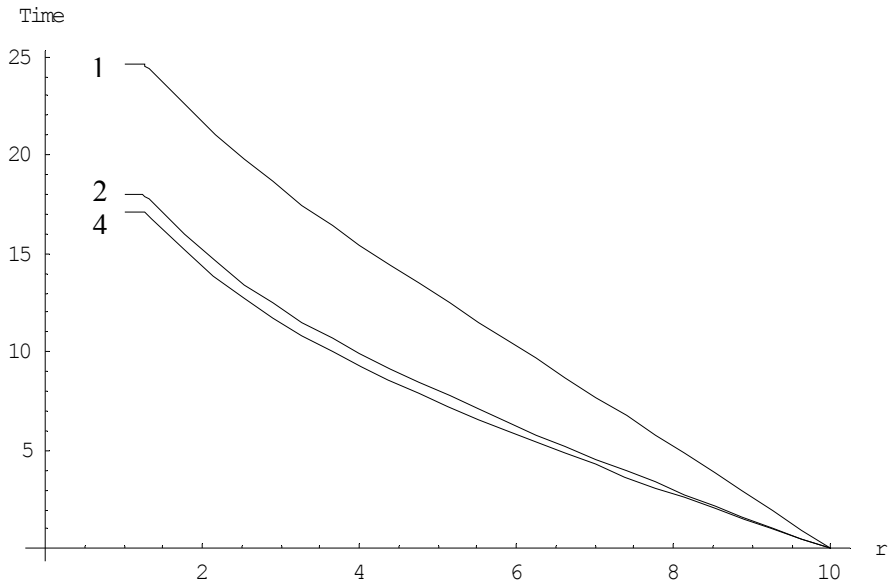

Graph of Time versus *r*, from *r* =10 to 1, along a radial geodesic for *E* = 1, 2 and 4 assuming that the test particles are travelling from infinity.

As can be seen, the last portion of the journey takes only an imperceptible amount of coordinate time.

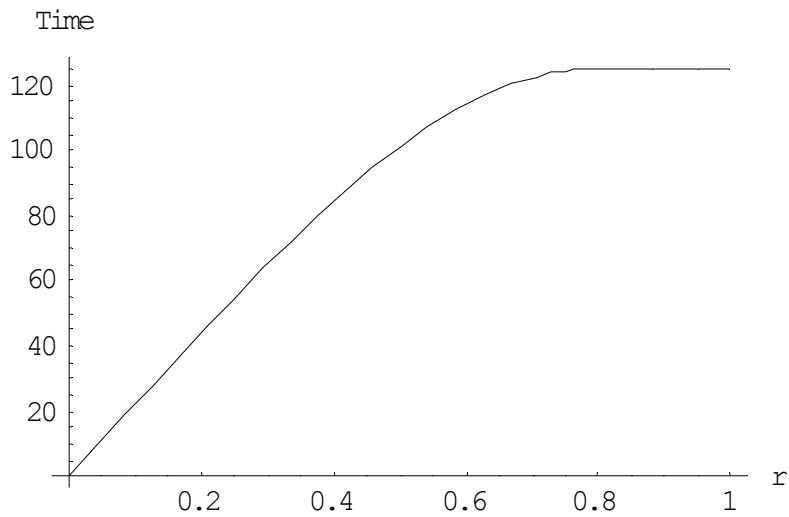

Graph of Time versus radial distance *r* from *r* = 0 to periphery of the ring along a radial geodesic for *E* = 1.00001

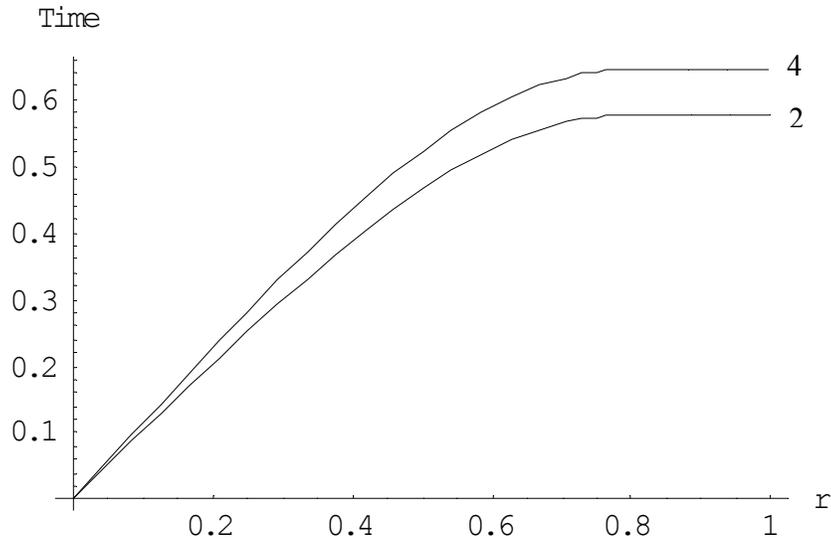

Graph of Time versus radial distance *r* from *r* = 0 to periphery of the ring along a radial geodesic for *E* = 2 and 4.

## Speed along the normal geodesics and the $(t, z)$ curve

The maximum speed does not reach that of light. Ultimately the test particle comes to rest at the centre for *E*=1.

When $a = 1$
Speed along the normal geodesics is given by

$$\left(1 - E^{-2} e^{2U}\right)^{1/2}, \quad U = U(z)$$

For different values of *E*, ($a = 1$), we get:

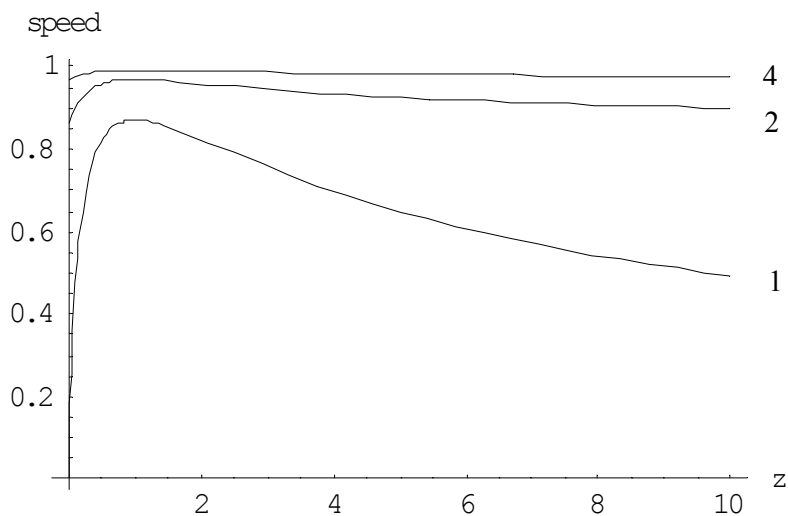

Graph of Speed versus normal distance *z*, from *z* = 10 to 0 along normal geodesics for values of *E* = 1, 2 and 4 assuming that the test particles are travelling from infinity.

When $z = 10$, $t = 0$ solving $t'[z] = -\dfrac{e^{\lambda - 2U}}{\sqrt{1 - E^{-2}e^{2U}}}$, the time taken to reach from $z = 10$ to $z = 0$ (centre) is as shown.

For different values of $E$, ($a = 1$), we get:

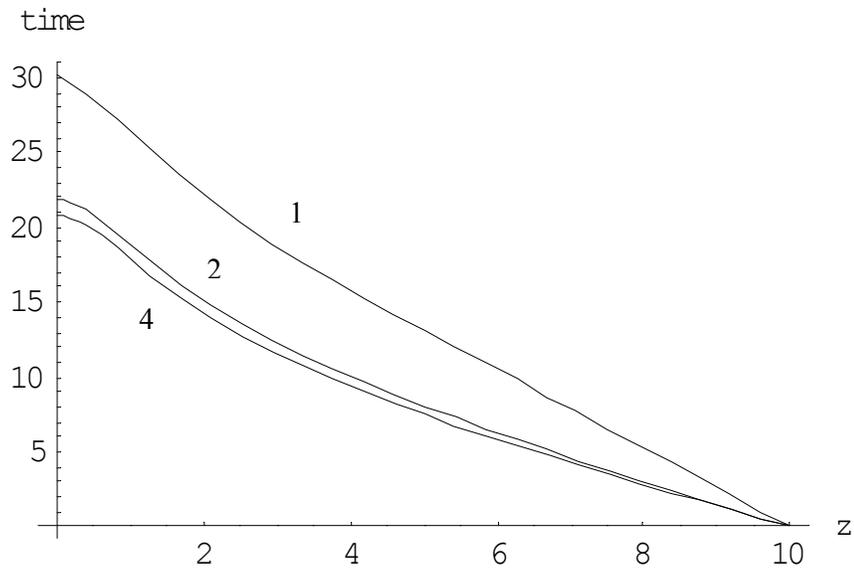

Graph of time versus normal distance $z$, from $z = 10$ to 0, along normal geodesics for $a = 1$, $E = 1, 2$ and 4. assuming that the test particles are travelling from infinity.

## Conclusion

The symmetric normal geodesics are well behaved. Inside the ring, from the origin ($r = 0$) to the periphery of the ring ($r = 1, z = 0$), the speed remains constant up to the ring because $U = 0$. It may be possible that there are geodesics threading the singularity.
 A more detailed evaluation of these geodesics might prove to be fruitful.

## Acknowledgements

Credit goes out to Priyanthi Wickramasuriya who advised us on the graphics for this paper.